\documentclass[11pt,a4paper]{article}
\usepackage{jheppub}
\usepackage[T1]{fontenc}

\allowdisplaybreaks

\newcommand{\0}{SO(10)}

\newcommand{\be}{\begin{equation}}
\newcommand{\ee}{\end{equation}}
\newcommand{\bea}{\begin{eqnarray}}
\newcommand{\eea}{\end{eqnarray}}

\newcommand{\phd}[1]{\Phi^{(D)}_{#1}}
\newcommand{\pht}[1]{\Phi^{(T)}_{#1}}

\title{Proton decay in a supersymmetric SO(10) model with missing partner mechanism}
\author{Lipei Du,}
\author{Xiaojia Li}
\author{and Da-Xin Zhang}
\affiliation{School of Physics
and State Key Laboratory of Nuclear Physics and Technology,\\
Peking University, Beijing 100871, China}
\emailAdd{lpdu@pku.edu.cn}
\emailAdd{shakalee@pku.edu.cn}
\emailAdd{dxzhang@pku.edu.cn}

\abstract{The extended supersymmetric  SO(10) model  with missing partner mechanism is studied.
An intermediate vacuum expectation value is incorporated which corresponds to the see-saw scale.
Gauge coupling unification is not broken explicitly.
Proton decay is found to satisfy the present experimental limits at the cost of
fine-tuning some parameters.}

\arxivnumber{1312.1786}
\begin{document}
\maketitle
\flushbottom
\pagenumbering{arabic}

\section{Introduction}

The supersymmetric version of SO(10) Grand Unified Theory (GUT) \cite{clark1982,aulakh1983,babu1993,bajc2003,goh2003,aulakh2004,fukuyama2004,bajc2004,fukuyama2005,fukuyama2005b,babu2005,aulakh2005,bertolini2006,aulakh2006,dutta2008,aulakh2009,aulakh2012,aulakh2013} is an interesting candidate for the new physics
beyond the Standard Model (SM). First, the fermions in every of the three generations are contained
in the same representation 16 of SO(10). These 16plets contain the right-handed neutrinos so that
neutrino masses and mixings can be taken into account through the see-saw mechanism.
Second, by including different Higgs multiplets,
fermion masses and mixings can be described correctly.
Third, once the two Higgs doublets for the charged fermion masses
are arranged at the electro-weak scale, supersymmetry (SUSY) protects their smallness against quadratic divergences.

The supersymmetric SO(10) GUT has also problems to be solved. The contents of the Higgs multiplets are not fixed {\it a priori}.
In the minimal version, the Higgs sector contains
10, 126, $\overline{126}$ and 210 \cite{aulakh2004,fukuyama2004,bajc2004,fukuyama2005,fukuyama2005b}. This brings in several difficulties in the predictions. The model is
highly constrained by the requirement of SUSY  at high energy scales. To account for the neutrino masses
and mixings, an intermediate scale given by the VEVs of $\overline{126}$ and $126$ is needed which, however,
breaks badly the unification of the gauge couplings.
Also, there is the so-called  doublet-triplet splitting (DTS) problem to  explain that why
the two Higgs doublets are light in the Minimal Supersymmetric SM (MSSM)
while all the other multiplets, especially the triplets in the same minimal representation 10, are all
superheavy.

The Missing Partner Mechanism (MPM) has been successfully used in supersymmetric GUT of  SU(5)
 \cite{MPM masiero1982,MPM grinstein1982,MPM hisano1995,MPM zheng2012,MPM zhang2012}
to  provide a natural realization of the  DTS within the Higgs multiplets.
This mechanism is then used in SO(10) model  \cite{pulido1983,ksbabu}. To account for the charged fermion masses,
this Minimal SO(10) model with MPM (Minimal SO10MPM) \cite{ksbabu} needs to be extended to include  more
$\overline{126}$ and $126$ while keeping the MPM  \cite{babu2}. However,
there are still difficulties to overcome, including how to incorporate  the see-saw mechanism and how to suppress
proton decay  in the model.

In the present work, we will study the extended SO10MPM.
The SO(10) symmetry is broken directly to the SM gauge group,
while an intermediate VEV to account for the see-saw scale is used which is only a small correction to the SO(10) symmetry breaking.
The other $126-\overline{126}$ need to have large VEVs not only to keep unification,
but also to suppress proton decay required by the experimental data.

In section \ref{two} we will review the Minimal SO10MPM briefly. In section \ref{three} the extended SO10MPM is constructed
and solved explicitly. Mass matrices of doublets and triplets are shown in section \ref{four}. Fermion masses and proton decay are studied in section \ref{five}. In section \ref{six} we will summarize.

\section{Review of the minimal SO10MPM}\label{two}

In the SU(5) models, the MPM is realized by introducing a U(1) symmetry to forbid the $5-\overline{5}$ to have direct mass term. By coupling  $5(\overline{5})$ Higgs to $\overline{50} (50)$, which contains a color triplet (anti-triplet) but no weak doublet,  through $75$ which breaks SU(5), the
color triplets in $5-\overline{5}$ gain GUT scale masses while the weak doublets are still massless.

In SO(10), the $5$ and $\overline{5}$ of SU(5)  constitute a 10, and the  $\overline{50}$ and $50$ of SU(5) are contained in  $126$ and $\overline{126}$, respectively, as
\begin{eqnarray}
126&=& 1+\overline{5}+10+\overline{15}+45+\overline{50},  \label{de126} \nonumber \\
\overline{126}&=& 1+5+\overline{10}+{15}+\overline{45}+{50}. \label{de126bar}
\end{eqnarray}
In total, a pair of $126-\overline{126}$ multiplets contains 2 pairs of doublets and 3 pairs of triplets.
In addition, the SU(5)-breaking $75$ is contained in $210$ of SO(10) which breaks SO(10) symmetry.
Then, in principle, the MPM of SU(5) can be embedded  into SO(10).
This needs to include more
Higgs in the model and to impose an extra U(1) symmetry to eliminate unwanted bilinear and trilinear couplings
from the superpotential, leaving some elements of the mass matrices to be zero.
Denoting the multiplets which cannot(can) get mass through couplings with itself or its conjugate as ``$light$''(``$heavy$'') fields,
the  Higgs fields
$${10}\times 2+{120}+{126/\overline{126}}+{210}$$
 are required in the minimal version \cite{ksbabu}, in which the $10$s and $120$ are $light$ with $Q=1$, $126-\overline{126}$ are $heavy$ with $Q=-1$, and $210$ is also $heavy$ with $Q=0$. The heavy fields $126-\overline{126}$ get their masses through couplings with
a singlet $X$ with $Q=2$, while the light fields get masses only through couplings with $126-\overline{126}$ and $210$. Fermions are contained in the three $16$plets with $Q=-\frac{1}{2}$, coupling only with the $light$ fields. It is easy to count that there are 4 (4) pairs of $light$ doublets (triplets), and 3 (4) pairs of $heavy$ doublets (triplets).
The mass matrices are
\begin{equation}\label{dscheme}
M_D=\bordermatrix{  & {D}_{light} &{D}_{heavy} \cr
\overline{D}_{light} & 0_{4\times 4} & O(G)_{4\times 3}\cr
\overline{D}_{heavy} & O(G)_{3\times 4} & O(G)_{3\times 3}
}
\end{equation}
for the doublets, and
\begin{equation}\label{tscheme}
M_T=\bordermatrix{  & {T}_{light} &{T}_{heavy} \cr
\overline{T}_{light} & 0_{4\times 4} & O(G)_{4\times 4}\cr
\overline{T}_{heavy} & O(G)_{4\times 4} & O(G)_{4\times 4}
}
\end{equation}
for the triplets.
Here, $0_{4\times 4}$ stands for a ${4\times 4}$ null matrix, while
$O(G)$s mean matrices whose elements are of the GUT scale. Generally speaking, $M_D$ has one zero (light) eigenvalue, and $M_T$ has none. Because the determinant of $M_D$ contains $7!=5040$ {terms}, and each term inevitably has one factor that comes from the $4\times 4$ null matrix part, this makes the determinant of $M_D$ vanished. Since the heavy part of $M_T$ is larger than that of $M_D$, all eigenvalues of $M_T$ remain superheavy, and the DTS is thus achieved.

However, one cannot fit all fermion masses with only $10$ and $120$ fields, a $light$ $\overline{126}$ must be introduced to couple with the fermions. Meanwhile, the $heavy$ side of fields must be extended to keep the MPM \cite{babu2}. We will study this extended SO10MPM in the next section which was not solved explicitly in  \cite{babu2}.

\section{Extended SO10MPM with 3 pairs of $126-\overline{126}$}\label{three}

In the Minimal SO10MPM the see-saw mechanism is not included to explain the neutrino masses. In fact, SO(10) has a basic conflict between incorporating the see-saw mechanism to generate the neutrino masses and realizing the unification of gauge couplings. In the type-I see-saw models  \cite{typeone1,typeone2,typeone3,typeone4}, the right-handed neutrinos need to have masses at the intermediate scale of the order of $O(10^{14})$GeV \cite{nurscale2002,nurscale2004,nurscale2013}. Similarly, in the type-II see-saw models, a weak triplet at the intermediate scale is needed. This intermediate scale, however, breaks the gauge coupling unification badly
since fields at this scale will change dramatically the running behaviors of the coupling constants.

To incorporate the see-saw mechanism, we use the extended SO10MPM with three pairs of $126-\overline{126}$.
The GUT scale VEVs are assigned to the MSSM singlets of these $126-\overline{126}$ except that of the $\overline{126}$
which couples to fermions has an intermediate VEV. The SO(10) symmetry is thus directly broken to MSSM symmetry, so that gauge coupling unification can be maintained hopefully as all particles other than the MSSM Higgs doublets are
superheavy. All the three pairs of  $126-\overline{126}$ are needed to keep SUSY unbroken at high energy,
as the conditions for the F- and D-flatness  are satisfied.

The model includes the following set of states in Table \ref{Qnumbers}, and their U(1) quantum numbers are also
assigned to eliminate unwanted masses and couplings.
The U(1) contains an anomaly which is canceled through the Green-Schwarz mechanism \cite{green1984,dine1987,atick1987,dine19872}.
\begin{table}[t]
\begin{center}
\begin{tabular}{|c|c|c|c|c|c|c|c|c|c|}
\hline
 &$\Psi_{i}(16)$ & $H_1(10)$ &  $H_2(10)$ &  $D(120)$ & $\Delta/\overline{\Delta} $ & $\Delta_1/\overline{\Delta}_1$ & $\Delta_2/\overline{\Delta}_2$ &$\Phi(210)$ &$X(1)$\\
\hline
$Q$ &$-\frac{1}{2}$ &1 &1 &1 &1 &-1 &-1 &0  &2\\
\hline
\end{tabular}
\caption{SO(10) multiplets and their U(1) charges. $\Psi_{i}(i=1,2,3)$ represent three generations of fermions,
$\Delta/\overline{\Delta} $, $\Delta_1/\overline{\Delta}_1 $, $\Delta_2/\overline{\Delta}_2 $ are all $126/\overline{126}$ multiplets, and $X$ is {an} \0 singlet whose VEV breaks the U(1) symmetry. $H_{1,2}$ and $D$ are 10plets and 120plet, respectively. Higgs multiplets with $Q=1$ are $light$ while the others are $heavy$. }\label{Qnumbers}
\end{center}
\end{table}

The general renormalizable Higgs superpotential is given by
\begin{eqnarray}  \label{superp2}
W&=&\frac{1}{2}m_{\Phi}\Phi^2+{m_{\Delta}}_{i}\overline{\Delta}_i\Delta
+{\overline{m}_{\Delta}}_{i}\overline{\Delta}\Delta_i
+\lambda\Phi^3 + \eta_{ij}X\overline{\Delta}_i\Delta_j  \nonumber \\
&+&(\lambda_{i}\overline{\Delta}_i\Delta +\overline{\lambda}_{i}\overline{\Delta}\Delta_i)\Phi
+(\beta_{ij}\Delta_i+\overline{\beta}_{ij}\overline{\Delta}_i)H_j\Phi
+(\gamma_i\Delta_i+\overline{\gamma}_i\overline{\Delta}_i)D\Phi,
\end{eqnarray}
where $i,j=1,2$. Repeated indices stand for summation, as usually understood.
Following the notation of  \cite{tfukuyama}, SO(10) breaks down to the
MSSM when the MSSM singlets  get VEVs as
\begin{eqnarray}
\Phi _{1}=\langle \Phi (1,1,1)\rangle , &&\Phi _{2}=\langle \Phi
(15,1,1)\rangle ,\quad\quad \Phi _{3}=\langle \Phi (15,1,3)\rangle,  \nonumber \\
&&v_{(1,2)R}=\langle \Delta (\overline{10},1,3)\rangle ,\quad \overline{v}_{(1,2)R}%
=\langle \overline{\Delta} (10,1,3)\rangle.
\end{eqnarray}
Here the $SU(4)_C\times SU(2)_L \times SU(2)_R$ indices are used to specify different singlets of MSSM.

Inserting these VEVs into (\ref{superp2}), we get:
\begin{eqnarray}  \label{superp2VEV}
\langle W\rangle &=&\frac{1}{2}m_{\Phi}(\Phi_1^2+\Phi_2^2+\Phi_3^2)+{m_{\Delta}}_{i}\overline{v}_{iR}v_R
+{\overline{m}_{\Delta}}_{i}\overline{v}_Rv_{iR}
+\lambda(\frac{1}{9\sqrt{2}}\Phi _{2}^{3}+\frac{1}{2%
\sqrt{6}}\Phi _{1}\Phi _{3}^{2}+\frac{1}{3\sqrt{2}}\Phi _{2}\Phi _{3}^{2})  \nonumber \\
&+&(\lambda_{i}\overline{v}_{iR}v_R +\overline{\lambda}_{i}\overline{v}_Rv_{iR})\Phi_0+ \eta_{ij}X\overline{v}_{iR}v_{jR},
\end{eqnarray}
where we have defined
\begin{equation}\label{phi0}
\Phi_0\equiv \left[\Phi_1\frac{1}{10\sqrt6}+\Phi_2\frac{1}{10\sqrt2}+\Phi_3\frac{1}{%
10}\right]
\end{equation}
for further convenience.

To preserve SUSY at high energy, the F- and D-flatness conditions are required.
The D-flatness condition requires
\begin{equation}  \label{dterm}
|v_{R}|^2+|v_{1R}|^2+|v_{2R}|^2=|\overline{v}_{R}|^2+|\overline{v}_{1R}|^2+|\overline{v}_{2R}|^2,
\end{equation}
while the F-flatness conditions are
\begin{equation}
\left\{ \frac{\partial }{\partial \Phi _{1}},
\frac{\partial }{\partial \Phi_{2}},
\frac{\partial }{\partial \Phi _{3}},
\frac{\partial }{\partial v_{R}},%
\frac{\partial }{\partial v_{1R}},%
\frac{\partial }{\partial v_{2R}},%
\frac{\partial }{\partial \overline{v}_{R}},
\frac{\partial }{\partial \overline{v}_{1R}},
\frac{\partial }{\partial \overline{v}_{2R}},
\frac{\partial }{\partial X}\right\} \langle W\rangle =0,
\label{partiale}
\end{equation}%
which are explicitly
\begin{eqnarray}
0&=&m_{\Phi}\Phi_1+\frac{\lambda\Phi_3^2}{2\sqrt6}+\frac{1}{10\sqrt6}(\lambda_{i}\overline{v}_{iR}v_R +\overline{\lambda}_{i}\overline{v}_Rv_{iR}),  \label{equphi1} \\
0&=&m_{\Phi}\Phi_2+\frac{\lambda\Phi_2^2}{3\sqrt2}+\frac{\lambda\Phi_3^2}{%
3\sqrt2}+\frac{1}{10\sqrt{2}}(\lambda_{i}\overline{v}_{iR}v_R +\overline{\lambda}_{i}\overline{v}_Rv_{iR}), \label{equphi2} \\
0&=&m_{\Phi}\Phi_3+\frac{\lambda\Phi_1\Phi_3}{\sqrt6}+\frac{%
\sqrt2\lambda\Phi_2\Phi_3}{3}+\frac{1}{10}(\lambda_{i}\overline{v}_{iR}v_R +\overline{\lambda}_{i}\overline{v}_Rv_{iR}),
\label{equphi3} \\
0&=&M_{1}\overline{v}_{1R}+M_{2}\overline{v}_{2R},                 \label{equvr} \\
0&=&\overline{M}_{1} \overline{v}_{R} +\eta_{i1}X\overline{v}_{iR}, \label{equvr1} \\
0&=&\overline{M}_{2} \overline{v}_{R} +\eta_{i2}X\overline{v}_{iR},   \label{equvr2} \\
0&=&\overline{M}_{1} v_{1R}+\overline{M}_{2} v_{2R},                \label{equvrbar}\\
0&=&M_{1}v_{R} +\eta_{1i}X{v}_{iR},                                  \label{equvr1bar} \\
0&=&M_{2}v_{R} +\eta_{2i}X{v}_{iR},                                  \label{equvr2bar} \\
0&=&\eta_{ij}\overline{v}_{iR}v_{jR}.                            \label{equX}
\end{eqnarray}
Here we have defined:
\begin{equation}  \label{Mij}
M_{i}\equiv {m_{\Delta}}_{i}+\lambda_{i}\Phi_0,\qquad
\overline{M}_{i}\equiv {\overline{m}_{\Delta}}_{i}+\overline{\lambda}_{i}\Phi_0.
\end{equation}

Now that there are 10 equations for the F-flatness and 1 for the D-flatness in total, but only 9 of these constraints are independent. The situation here is comparable with that of the minimal SUSY SO(10) model in  \cite{aulakh2004,bajc2004}, where there are 5 equations for the F-flatness and 1 for the D-flatness, but only 5 are independent.
We can solve the above equations as follows. The first three equations in the F-flatness conditions,  (\ref{equphi1}--\ref{equphi3}), are solved in the same way as in the  minimal SUSY SO(10) model which gives
\begin{eqnarray}
\Phi_1&=&-\frac{\sqrt6 m_{\Phi}}{\lambda}\frac{x(1-5x^2)}{(1-x)^2},    \label{solphi1} \\
\Phi_2&=&-\frac{3\sqrt2 m_{\Phi}}{\lambda}\frac{(1-2x-x^2)}{(1-x)},    \label{solphi2} \\
\Phi_3&=&\frac{6 m_{\Phi}}{\lambda}x,    \label{solphi3} \\
(\lambda_{i}\overline{v}_{iR}v_R+\overline{\lambda}_{i}\overline{v}_Rv_{iR})&=&\frac{60m_{\Phi}^2}{\lambda}\frac{%
x(1-3x)(1+x^2)}{(1-x)^2}.  \label{solvrvrbar}
\end{eqnarray}
For $x$ is not to be taken special values to generate accidental intermediate symmetries,
all the VEVs $\Phi_{1,2,3}$ are at GUT scale  \cite{aulakh2004,bajc2004}.

The next three equations, (\ref{equvr}--\ref{equvr2}), are linear homogeneous equations about $\overline{v}_{R}$, $\overline{v}_{1R}$ and $\overline{v}_{2R}$, so they have non-zero solutions only when
\begin{equation}
\textrm{det}
\ \left(\begin{array}{ccc} \label{detvr}
     0&M_1 & ~M_2 \\
     \overline{M}_{1}&\eta_{11}X &\eta_{21}X \\
     \overline{M}_{2}&\eta_{12}X &\eta_{22}X
     \end{array}\right)=0.
\end{equation}
Then, among $\overline{v}_{R}$, $\overline{v}_{1R}$ and $\overline{v}_{2R}$, only one can be considered as free which is chosen to be $\overline{v}_{1R}$ without loss of generality.
Similarly, equations (\ref{equvrbar}--\ref{equvr2bar}) require
\begin{equation}
\textrm{det}
\ \left(\begin{array}{ccc} \label{detvrbar}
      0&\overline{M}_{1} & ~\overline{M}_{2} \\
     {M}_{1}&\eta_{11}X &\eta_{12}X \\
     {M}_{2}&\eta_{21}X &\eta_{22}X
     \end{array}\right)=0
\end{equation}
to have non-zero solutions for $v_{R}$, $v_{1R}$ and $v_{2R}$, and
${v}_{1R}$ is taken to be free among them.
Note that the condition (\ref{detvrbar}) is exactly the same as (\ref{detvr}).
However, the transpose relation of the matrices in (\ref{detvr}) and in (\ref{detvrbar}) means that
the two sets of equations (\ref{equvr}--\ref{equvr2}) and (\ref{equvrbar}--\ref{equvr2bar}) have different
solutions.
The last equation,  (\ref{equX}), holds automatically  after taking (\ref{equphi1}--\ref{equvr2bar}) into account.

When all the superpotential parameters are fixed,
this condition (\ref{detvr}) or (\ref{detvrbar}) determines the value of $\Phi_0$, and thus $x$ can be extracted from
(\ref{phi0}) and (\ref{solphi1}--\ref{solphi3}).
The ratio of $\overline{v}_{1R}$ to ${v}_{1R}$  is determined by the D-flatness condition (\ref{dterm}).
With $x$ being solved already, all these VEVs of $126-\overline{126}$ are given.

To realize the see-saw mechanism,
an intermediate value for $\overline{v}_{R}$ is taken. All the other VEVs are taken at the GUT scale.
Consequently, there is no Higgs existing at the intermediate scale. The only exceptions are the right-handed
neutrinos needed in the type-I see-saw.

\section{Mass matrices for doublets and triplets}\label{four}

In the Extended SO10MPM model, there are  6 (7) pairs of $light$ doublets (triplets) and 5 (7) pairs of $heavy$ doublets (triplets).
The mass matrix for the doublets is written as
\begin{equation}
M_D=\left(
\begin{array}{cc}
A_{11(6\times 6)} &A_{12(6\times 5)} \\
A_{21(5\times 6)}&A_{22(5\times 5)}
\end{array}
\right),
\end{equation}
where the bases are
$$ H_{1(1,2,2)}^{(1,2,\frac{1}{2})},H_{2(1,2,2)}^{(1,2,\frac{1}{2})},
D_{(1,2,2)}^{(1,2,\frac{1}{2})},D_{(15,2,2)}^{(1,2,\frac{1}{2})}
,\Delta_{(15,2,2)}^{(1,2,\frac{1}{2})}
,\overline{\Delta}_{(15,2,2)}^{(1,2,\frac{1}{2})}$$
for the first 6 columns,
$$ H_{1(1,2,2)}^{(1,2,-\frac{1}{2})},H_{2(1,2,2)}^{(1,2,-\frac{1}{2})},
D_{(1,2,2)}^{(1,2,-\frac{1}{2})},D_{(15,2,2)}^{(1,2,-\frac{1}{2})}
,\overline{\Delta}_{(15,2,2)}^{(1,2,-\frac{1}{2})}
,{\Delta}_{(15,2,2)}^{(1,2,-\frac{1}{2})}$$
for the first 6 rows,
$$ \Delta_{1(15,2,2)}^{(1,2,\frac{1}{2})}
,\overline{\Delta}_{1(15,2,2)}^{(1,2,\frac{1}{2})},\Delta_{2(15,2,2)}^{(1,2,\frac{1}{2})}
,\overline{\Delta}_{2(15,2,2)}^{(1,2,\frac{1}{2})}
, \Phi_{(\overline{10},2,2)}^{(1,2,\frac{1}{2})}$$
for the last 5 columns, and
$$ \overline{\Delta}_{1(15,2,2)}^{(1,2,-\frac{1}{2})}
,{\Delta}_{1(15,2,2)}^{(1,2,-\frac{1}{2})},\overline{\Delta}_{2(15,2,2)}^{(1,2,-\frac{1}{2})}
,{\Delta}_{2(15,2,2)}^{(1,2,-\frac{1}{2})}
, \Phi_{({10},2,2)}^{(1,2,-\frac{1}{2})}$$
for the last 5 rows.
In these bases, the subscripts and the superscripts label the $SU(4)_C \times SU(2)_L \times SU(2)_R$ and
the SM representations, respectively.
In the $M_D$, $A_{11}$ is a $6\times 6$ null matrix, while
\begin{equation}\label{MD12}
A_{12}=\left(
\begin{array}{cccccc}
  \beta_{11}\phd{H\Delta} & \overline{\beta}_{11}\phd{H\overline{\Delta}} & {\beta}_{21}\phd{H\Delta} & \overline{\beta}_{21}\phd{H\overline{\Delta}} & -\frac{\overline{\beta}_{i1} \overline{v}_{iR}}{\sqrt{5}} \\
  {\beta}_{12}\phd{H\Delta} & \overline{\beta}_{12}\phd{H\overline{\Delta}} & {\beta}_{22}\phd{H\Delta} & \overline{\beta}_{22}\phd{H\overline{\Delta}} & -\frac{\overline{\beta}_{i2} \overline{v}_{iR}}{\sqrt{5}} \\
  \frac{{\gamma}_{1} \Phi_{3}}{4\sqrt{30}} & \frac{\overline{\gamma}_{1} \Phi_{3}}{4\sqrt{30}} & \frac{{\gamma}_{2} \Phi_{3}}{4\sqrt{30}} & \frac{\overline{\gamma}_{2} \Phi_{3}}{4\sqrt{30}} & -\frac{\overline{\gamma}_{i} \overline{v}_{iR}}{2\sqrt{30}} \\
 {\gamma}_{1}\phd{D\Delta} & \overline{\gamma}_{1}\phd{D\overline{\Delta}} & {\gamma}_{2}\phd{D\Delta} & \overline{\gamma}_{2}\phd{D\overline{\Delta}} & -\frac{\overline{\gamma}_{i} \overline{v}_{iR}}{2\sqrt{10}} \\
  \overline{m}_{{\Delta}_{1}}+\overline{\lambda}_{1}\phd{\overline{\Delta}\Delta} & 0 & \overline{m}_{{\Delta}_{2}}+\overline{\lambda}_{2}\phd{\overline{\Delta}\Delta} & 0 & 0 \\
  0& m_{\Delta_{1}}+{\lambda}_{1}\phd{\Delta\overline{\Delta}} & 0 & m_{\Delta_{2}}+{\lambda}_{2}\phd{\Delta\overline{\Delta}}  & \frac{{\lambda}_{i} \overline{v}_{iR}}{10} \\
\end{array}
\right),
\end{equation}
\\
\begin{equation}\label{MD21}
A_{21}=\left(
\begin{array}{cccccc}
 \overline{\beta}_{11}\phd{H\Delta} & \overline{\beta}_{12}\phd{H\Delta} & \frac{\overline{\gamma}_{1} \Phi_{3}}{4\sqrt{30}} & \overline{\gamma}_{1}\phd{D\Delta} & m_{\Delta_{1}}+{\lambda}_{1}\phd{\Delta\overline{\Delta}} & 0  \\
 \beta_{11}\phd{H\overline{\Delta}} & {\beta}_{12}\phd{H\overline{\Delta}} & \frac{{\gamma}_{1} \Phi_{3}}{4\sqrt{30}} & {\gamma}_{1}\phd{D\overline{\Delta}} & 0 & \overline{m}_{{\Delta}_{1}}+\overline{\lambda}_{1}\phd{\overline{\Delta}\Delta}  \\
 \overline{\beta}_{21}\phd{H\Delta} & \overline{\beta}_{22}\phd{H\Delta} & \frac{\overline{\gamma}_{2} \Phi_{3}}{4\sqrt{30}} & \overline{\gamma}_{2}\phd{D\Delta} & m_{\Delta_{2}}+{\lambda}_{2}\phd{\Delta\overline{\Delta}} & 0  \\
 {\beta}_{21}\phd{H\overline{\Delta}} & {\beta}_{22}\phd{H\overline{\Delta}} & \frac{{\gamma}_{2} \Phi_{3}}{4\sqrt{30}} & {\gamma}_{2}\phd{D\overline{\Delta}} & 0 & \overline{m}_{{\Delta}_{2}}+\overline{\lambda}_{2}\phd{\overline{\Delta}\Delta}   \\
 -\frac{\beta_{i1} {v}_{iR}}{\sqrt{5}} & -\frac{{\beta}_{i2} {v}_{iR}}{\sqrt{5}} & -\frac{{\gamma}_{i} {v}_{iR}}{2\sqrt{30}} & -\frac{{\gamma}_{i} {v}_{iR}}{2\sqrt{10}} & 0 & \frac{\overline{\lambda}_i {v}_{iR}}{10}
\end{array}
\right),
\end{equation}
and
\begin{equation}\label{MD22}
A_{22}=\left(
\begin{array}{ccccc}
 {\eta}_{11} X & 0 & {\eta}_{12} X & 0 & 0 \\
 0 & {\eta}_{11} X & 0 & {\eta}_{21} X & \frac{\overline{\lambda}_1\overline{v}_{R}}{10} \\
 {\eta}_{21} X & 0 & {\eta}_{22} X & 0 & 0 \\
 0 & {\eta}_{12} X & 0 & {\eta}_{22} X & \frac{\overline{\lambda}_2 \overline{v}_{R}}{10} \\
 0 & \frac{{\lambda}_{1} {v}_{R}}{10} & 0 & \frac{{\lambda}_{2} {v}_{R}}{10} & m_{\Phi}+\frac{\lambda \Phi_{2}}{\sqrt{2}}+\frac{\lambda \Phi_{3}}{2}
\end{array}
\right),
\end{equation}
where for simplicity we have defined
\begin{equation}\label{doubelements}
\begin{split}
\phd{H\Delta}&= \frac{\Phi_{2}}{\sqrt{10}}-\frac{\Phi_{3}}{2\sqrt{5}}, \qquad \qquad\qquad\qquad
\phd{H\overline{\Delta}} = -\frac{\Phi_{2}}{\sqrt{10}}-\frac{\Phi_{3}}{2\sqrt{5}},\\
\phd{D\Delta}&= \frac{\Phi_{1}}{4\sqrt{15}}-\frac{\Phi_{3}}{6\sqrt{10}}, \qquad\qquad\qquad\quad
\phd{D\overline{\Delta}}= \frac{\Phi_{1}}{4\sqrt{15}}+\frac{\Phi_{3}}{6\sqrt{10}},\\
\phd{\Delta\overline{\Delta}}&=\frac{\Phi_{2}}{15\sqrt{2}}-\frac{\Phi_{3}}{30}, \qquad\qquad\qquad\quad\quad
\phd{\overline{\Delta}\Delta}=\frac{\Phi_{2}}{15\sqrt{2}}+\frac{\Phi_{3}}{30}.\\
\end{split}
\end{equation}

It is easy to see that there is one zero-valued eigen-state in the row and another one in the column, which
are the MSSM Higgs doublets.
It can be also noted that these light doublets contain no components from the doublets in $210$ which break
$B-L$ quantum numbers. Consequently, the $SU(2)_L$ triplets in $\overline{126}$s has no
VEV, which excludes the type-II see-saw mechanism \cite{seesawII} in the model.

The mass matrix for the Higgs triplets are divided into four $7\times 7$ blocks as
\begin{equation}\label{triplet}
M_T=\left(
\begin{array}{cc}
B_{11(7\times 7)} &B_{12(7\times 7)} \\
B_{21(7\times 7)}& B_{22(7\times 7)}
\end{array}
\right),
\end{equation}
where the bases are
$$ H_{1(6,1,1)}^{(3,1,-\frac{1}{3})},H_{2(6,1,1)}^{(3,1,-\frac{1}{3})},
D_{(6,1,3)}^{(3,1,-\frac{1}{3})},D_{(10,1,1)}^{(3,1,-\frac{1}{3})}
,\overline{\Delta}_{(6,1,1)}^{(3,1,-\frac{1}{3})},
\overline{\Delta}_{(10,1,3)}^{(3,1,-\frac{1}{3})},\Delta_{(6,1,1)}^{(3,1,-\frac{1}{3})}$$
for the first 7 columns,
$$ H_{1(6,1,1)}^{(\overline{3},1,\frac{1}{3})},H_{2(6,1,1)}^{(\overline{3},1,\frac{1}{3})},
D_{(6,1,3)}^{(\overline{3},1,\frac{1}{3})},D_{(\overline{10},1,1)}^{(\overline{3},1,\frac{1}{3})}
,\overline{\Delta}_{(6,1,1)}^{(\overline{3},1,\frac{1}{3})}
,{\Delta}_{(6,1,1)}^{(\overline{3},1,\frac{1}{3})},{\Delta}_{(\overline{10},1,3)}^{(\overline{3},1,\frac{1}{3})}$$
for the first 7 rows,
$$ \Delta_{1(6,1,1)}^{(3,1,-\frac{1}{3})},\overline{\Delta}_{1(6,1,1)}^{(3,1,-\frac{1}{3})},
\overline{\Delta}_{1(10,1,3)}^{(3,1,-\frac{1}{3})},\Delta_{2(6,1,1)}^{(3,1,-\frac{1}{3})},
\overline{\Delta}_{2(6,1,1)}^{(3,1,-\frac{1}{3})},
\overline{\Delta}_{2(10,1,3)}^{(3,1,-\frac{1}{3})},
\Phi_{(15,1,3)}^{(3,1,-\frac{1}{3})}$$
for the last 7 columns, and
$$ \overline{\Delta}_{1(6,1,1)}^{(\overline{3},1,\frac{1}{3})}
,{\Delta}_{1(6,1,1)}^{(\overline{3},1,\frac{1}{3})},{\Delta}_{1(\overline{10},1,3)}^{(\overline{3},1,\frac{1}{3})}
,\overline{\Delta}_{2(6,1,1)}^{(\overline{3},1,\frac{1}{3})}
,{\Delta}_{2(6,1,1)}^{(\overline{3},1,\frac{1}{3})},{\Delta}_{2(\overline{10},1,3)}^{(\overline{3},1,\frac{1}{3})},
\Phi_{(15,1,3)}^{(\overline{3},1,\frac{1}{3})}$$
for the last 7 rows.
In the $M_T$, $B_{11}$ is a $7\times 7$ null matrix, the rest are
\begin{equation}\label{MT12}
B_{12}=\left(
\begin{array}{ccccccc}
 \beta_{11}\pht{H\Delta} & \overline{\beta}_{11}\pht{H\overline{\Delta}} & -\overline{\beta}_{11}\frac{\sqrt{2}\Phi_{3}}{\sqrt{15}} & \beta_{21}\pht{H\Delta} & \overline{\beta}_{21}\pht{H\overline{\Delta}} & -\overline{\beta}_{21}\frac{\sqrt{2}\Phi_{3}}{\sqrt{15}} & \frac{\overline{\beta}_{i1}\overline{v}_{iR}}{\sqrt{5}}  \\
 \beta_{12}\pht{H\Delta} & \overline{\beta}_{12}\pht{H\overline{\Delta}} & -\overline{\beta}_{12}\frac{\sqrt{2}\Phi_{3}}{\sqrt{15}} & \beta_{22}\pht{H\Delta} & \overline{\beta}_{22}\pht{H\overline{\Delta}} & -\overline{\beta}_{22}\frac{\sqrt{2}\Phi_{3}}{\sqrt{15}} & \frac{\overline{\beta}_{i2}\overline{v}_{iR}}{\sqrt{5}}  \\
  \frac{\gamma_{1}\Phi_{3}}{12\sqrt{5}} & \frac{\overline{\gamma}_{1}\Phi_{3}}{12\sqrt{5}} & \frac{\overline{\gamma}_{1}\Phi_{2}}{6\sqrt{5}} & \frac{\gamma_{2}\Phi_{3}}{12\sqrt{5}} & \frac{\overline{\gamma}_{2}\Phi_{3}}{12\sqrt{5}} & \frac{\overline{\gamma}_{2}\Phi_{2}}{6\sqrt{5}} & \frac{\overline{\gamma}_{i}\overline{v}_{iR}}{2\sqrt{15}}  \\
  -\frac{\gamma_{1}\Phi_{2}}{6\sqrt{10}} & \frac{\overline{\gamma}_{1}\Phi_{2}}{6\sqrt{10}} & \frac{\overline{\gamma}_{1}\Phi_{3}}{6\sqrt{10}} & -\frac{\gamma_{2}\Phi_{2}}{6\sqrt{10}} & \frac{\overline{\gamma}_{2}\Phi_{2}}{6\sqrt{10}} & \frac{\overline{\gamma}_{2}\Phi_{3}}{6\sqrt{10}} & \frac{\overline{\gamma}_{i}\overline{v}_{iR}}{2\sqrt{15}}  \\
 \overline{m}_{\Delta_1} & 0 & 0 & \overline{m}_{\Delta_2} & 0 & 0 & 0   \\
  0 & {m}_{\Delta_1} & \frac{\lambda_1 \Phi_3}{15\sqrt{2}} & 0 & {m}_{\Delta_2} & \frac{\lambda_2 \Phi_3}{15\sqrt{2}} &  -\frac{{\lambda}_{i}\overline{v}_{iR}}{10\sqrt{3}}  \\
 0 & \frac{\lambda_1 \Phi_3}{15\sqrt{2}} & {m}_{\Delta_1}+\lambda_1 \pht{\Delta} & 0 &\frac{\lambda_2 \Phi_3}{15\sqrt{2}} & {m}_{\Delta_2}+\lambda_2 \pht{\Delta} & -\frac{{\lambda}_{i}\overline{v}_{iR}}{5\sqrt{6}}
\end{array}
\right),
\end{equation}
\\
\begin{equation}\label{MT21}
B_{21}=\left(
\begin{array}{ccccccc}
\overline{\beta}_{11}\pht{H\Delta} & \overline{\beta}_{12}\pht{H\Delta} & \frac{\overline{\gamma}_{1}\Phi_{3}}{12\sqrt{5}} & -\frac{\overline{\gamma}_{1}\Phi_{2}}{6\sqrt{10}} & 0 & 0 & {m}_{\Delta_1}  \\
{\beta}_{11}\pht{H\overline{\Delta}} & {\beta}_{12}\pht{H\overline{\Delta}}& \frac{{\gamma}_{1}\Phi_{3}}{12\sqrt{5}} & \frac{{\gamma}_{1}\Phi_{2}}{6\sqrt{10}}& \overline{m}_{\Delta_1} & \frac{\overline{\lambda}_{1}\Phi_{3}}{15\sqrt{2}} & 0 \\
-{\beta}_{11}\frac{\sqrt{2}\Phi_{3}}{\sqrt{15}} & -{\beta}_{12}\frac{\sqrt{2}\Phi_{3}}{\sqrt{15}} &  \frac{{\gamma}_{1}\Phi_{2}}{6\sqrt{5}} & \frac{{\gamma}_{1}\Phi_{3}}{6\sqrt{10}} & \frac{\overline{\lambda}_1 \Phi_3}{15\sqrt{2}} & \overline{m}_{\Delta_1}+\overline{\lambda}_1 \pht{\Delta} & 0  \\
  \overline{\beta}_{21}\pht{H\Delta} & \overline{\beta}_{22}\pht{H\Delta} & \frac{\overline{\gamma}_{2}\Phi_{3}}{12\sqrt{5}} &  -\frac{\overline{\gamma}_{2}\Phi_{2}}{6\sqrt{10}} & 0 & 0 &{m}_{\Delta_2}  \\
 {\beta}_{21}\pht{H\overline{\Delta}} & {\beta}_{22}\pht{H\overline{\Delta}} &  \frac{{\gamma}_{2}\Phi_{3}}{12\sqrt{5}} & \frac{{\gamma}_{2}\Phi_{2}}{6\sqrt{10}} & \overline{m}_{\Delta_2} & \frac{\overline{\lambda}_2 \Phi_3}{15\sqrt{2}} & 0   \\
 -{\beta}_{21}\frac{\sqrt{2}\Phi_{3}}{\sqrt{15}} &  -{\beta}_{22}\frac{\sqrt{2}\Phi_{3}}{\sqrt{15}} & \frac{{\gamma}_{2}\Phi_{2}}{6\sqrt{5}} & \frac{{\gamma}_{2}\Phi_{3}}{6\sqrt{10}} & \frac{\overline{\lambda}_2 \Phi_3}{15\sqrt{2}} & \overline{m}_{\Delta_2}+\overline{\lambda}_2 \pht{\Delta} & 0  \\
 \frac{{\beta}_{i1}{v}_{iR}}{\sqrt{5}} & \frac{{\beta}_{i2}{v}_{iR}}{\sqrt{5}} & \frac{{\gamma}_{i}{v}_{iR}}{2\sqrt{15}} & \frac{{\gamma}_{i}{v}_{iR}}{2\sqrt{15}} & -\frac{\overline{\lambda}_{i}{v}_{iR}}{10\sqrt{3}} & -\frac{\overline{\lambda}_{i}{v}_{iR}}{5\sqrt{6}} & 0
\end{array}
\right),
\end{equation}
and
\begin{equation}\label{MT22}
B_{22}=\left(
\begin{array}{ccccccc}
 \eta_{11}X & 0 & 0 & \eta_{12}X & 0 & 0 & 0  \\
 0 & \eta_{11}X & 0 & 0 & \eta_{21}X & 0 & -\frac{\overline{\lambda}_{1}\overline{v}_{R}}{10\sqrt{3}}  \\
 0 & 0 & \eta_{11}X & 0 & 0 & \eta_{21}X & -\frac{\overline{\lambda}_{1}\overline{v}_{R}}{5\sqrt{6}}  \\
 \eta_{21}X& 0 & 0 & \eta_{22}X & 0 & 0 & 0  \\
 0 & \eta_{12}X & 0 & 0 & \eta_{22}X & 0 & -\frac{\overline{\lambda}_{2}\overline{v}_{R}}{10\sqrt{3}}   \\
  0 & 0 & \eta_{12}X & 0 & 0 & \eta_{22}X & -\frac{\overline{\lambda}_{2}\overline{v}_{R}}{5\sqrt{6}}  \\
 0 & -\frac{{\lambda}_{1}{v}_{R}}{10\sqrt{3}} & -\frac{\lambda_{1}{v}_{R}}{5\sqrt{6}} & 0 & -\frac{{\lambda}_{2}{v}_{R}}{10\sqrt{3}} & -\frac{{\lambda}_{2}{v}_{R}}{5\sqrt{6}} & m_{\Phi}+\lambda(\frac{\Phi_1}{\sqrt{6}}+\frac{\Phi_2}{3\sqrt{2}}+\frac{2\Phi_3}{3})   \\
\end{array}
\right),
\end{equation}
where
\begin{eqnarray}\label{doubelements}
\pht{H\Delta}&=&-\frac{\Phi_{1}}{\sqrt{10}}+\frac{\Phi_{2}}{\sqrt{30}},\nonumber\\
\pht{H\overline{\Delta}}&=&-\frac{\Phi_{1}}{\sqrt{10}}-\frac{\Phi_{2}}{\sqrt{30}},\\
\pht{\Delta}&=&\frac{\Phi_{1}}{10\sqrt{6}}+\frac{\Phi_{2}}{30\sqrt{2}}.\nonumber
\end{eqnarray}
We can observe that there is no zero eigenvalue of the triplet mass matrix. Also,
all the other Higgs are massive except those Goldstone modes. This justifies the realization of the MPM in the extended model.

\section{Fermion masses and proton decay}\label{five}

In the present model, the fermion sector is described by the superpotential
\begin{equation}
W=Y_{10_1}^{ij}\Psi_i\Psi_j H_1+Y_{10_2}^{ij}\Psi_i\Psi_j H_2+Y_{120}^{ij}\Psi_i\Psi_j D+Y_{126}^{ij}\Psi_i\Psi_j \overline{\Delta}.
\end{equation}
In  \cite{bertolini,Patel} the fermion masses can be fitted by using only one $10_H$ and one $\overline{126}_H$.
As in \cite{bertolini} for example, the resultant fermion masses are in accord with their experimental values except the electron mass.
In present, although all the Higgs $H_{1,2}$, $D$ and $\overline{\Delta}$ contribute to the fermion masses,
we can also use the $H_{1}$ and $\overline{\Delta}$ as in   \cite{bertolini}, taking
contributions from $H_{2}$ and $D$ as small corrections to the electron mass. These corrections are
suppressed by a factor of
$$\frac{Y_{10_2}^{ij}}{Y_{126}^{ij}} \sim \frac{Y_{120}^{ij}}{Y_{126}^{ij}} \sim \frac{m_e}{m_\tau} \sim 10^{-4}$$
and are negligible in studying proton decay.

At the GUT scale the fermion masses are taken from  \cite{C.R.Das} for $\tan\beta(M_{SUSY})=10$, $\mu=2\times 10^{16} \textrm{GeV}$. After numerical fitting, we get the Yukawa couplings $Y_{10_1}^{ij}$ and $Y_{126}^{ij}$ in the $u$-diagonal basis
\begin{equation}\label{Y10}
Y_{10_1}^{ij}=\left(
\begin{array}{ccc}
 {0.000281}&-0.000784-0.000103 i&0.00760 + 0.00270 i \\
 -0.000784-0.000103 i & 0.00174&-0.0337 + 0.000650 i \\
0.00760+0.00270 i&-0.0337 + 0.000650 i&0.999
\end{array}
\right),
\end{equation}
\\
\begin{equation}\label{Y126}
Y_{126}^{ij}=\left(
\begin{array}{ccc}
 0.0000651&-0.0000535+{7.06\times 10^{-6}} i&-0.000519 -0.000184i \\
-0.0000535+{7.06\times 10^{-6}} i&-0.00180&0.00231 -{4.45\times 10^{-5}}i \\
-0.000519 -0.000184i&0.00231 -{4.45\times 10^{-5}} i&-0.000958
\end{array}
\right).
\end{equation}
One can easily verify these couplings by calculating the mass eigenvalues of the following matrices
\begin{eqnarray}\label{doubelements}
M_u&=&(\alpha^u Y_{10_1}+\beta^u Y_{126}) v_u, \nonumber\\
M_d&=&(\alpha^d Y_{10_1}+\beta^d Y_{126}) v_d, \\
M_e&=&(\alpha^d Y_{10_1}-3\beta^d Y_{126}) v_d, \nonumber
\end{eqnarray}
where $v_u=123.8 \textrm{GeV}$ and $v_d=17.87 \textrm{GeV}$ (see Table 5 of  \cite{C.R.Das}). $\alpha^u=0.6647$, $\beta^u=-0.7471$, $\alpha^d=0.06816$ and $\beta^d=-0.9977$ come from numerical fitting. We also noticed that the
normalizations $(\alpha^u)^2+(\beta^u)^2=1$ and $(\alpha^d)^2+(\beta^d)^2=1$ are not accurate due to the existence of $H_2$ and $D(120)$, but the deviations are small to be neglected reasonably.

In studying proton decay via dimension-5 operators,
we limit our analysis to $LLLL$ operators only, although contributions from $RRRR$ operators are also sizable \cite{nath1985,goto1999}.
The resultant operators are contained in the superpotential
\begin{eqnarray}\label{w5}
W_5=C^{ijkl}(Q_i Q_j)(Q_k L_l),
\end{eqnarray}
where the contractions of the indices are understood as
\begin{eqnarray}
(Q_i Q_j)(Q_k L_l)=\epsilon_{\alpha\beta\gamma}(u_i^\alpha {d'}_j^{\beta}-{d'}_i^\alpha u_j^{\beta})(u_k^\gamma e_l-{d'}_k^\gamma \nu_l).
\end{eqnarray}
In (\ref{w5}) we have defined
\begin{eqnarray}\label{Cijkl}
C^{ijkl}&=&C_{11}^{ijkl}+C_{12}^{ijkl}+C_{21}^{ijkl}+C_{22}^{ijkl},\nonumber \\
C_{11}^{ijkl} &=&Y_{10_1}^{ij} (M_T^{-1})_{11} Y_{10_1}^{kl},\qquad
C_{12}^{ijkl} = Y_{10_1}^{ij} (M_T^{-1})_{15} Y_{126}^{kl},\\
C_{21}^{ijkl}&= &Y_{126}^{ij} (M_T^{-1})_{51} Y_{10_1}^{kl},\qquad
C_{22}^{ijkl}= Y_{126}^{ij} (M_T^{-1})_{55} Y_{126}^{kl},\nonumber
\end{eqnarray}
where $M_T^{-1}$ is the inverse of the triplet mass matrix $M_T$ in (\ref{triplet}),
and only
$(M_T^{-1})_{11}$, $(M_T^{-1})_{15}$, $(M_T^{-1})_{51}$ and $(M_T^{-1})_{55}$ contribute to
the  $LLLL$-type proton decay
because of the  $SU(4)_C\times SU(2)_L\times SU(2)_R$ basis in the triplet mass matrix.
The relevant up-left block in $M_T^{-1}$ equals to inverse of the effective triplet mass matrix
which is got by integrating out the down-right block $B_{22}$
in (\ref{triplet}),
\begin{equation}\label{effectivemass}
 M_{eff}=-B_{12}\cdot B_{22}^{-1} \cdot B_{21}.
 \end{equation}

Running the dimension-5 operators down to the SUSY scale and dressing them by wino-loops, we get the
 four-fermion operators. For the dominant decay modes $p\rightarrow K^+\nu_l$, the coefficients are
\begin{eqnarray}
C_{sudv_l}&=&2\ {\frac{ {\alpha_2}}{2 {\pi}}f_{\Delta}} \  {A_L} \  {A_S} \ \left( \sum_{j,k=1}^3 C^{1jkl} {V}_{j2} {V}_{k1}+ {\sum_{i,j=1}^3}C^{ij1l} {V}_{j2} {V}_{i1}\right),\\
C_{dusv_l}&=&2\ {\frac{ {\alpha_2}}{2 {\pi}}f_{\Delta}} \ {A_L}\ {A_S}\ \left( {\sum_{j,k=1}^3}C^{1jkl} {V}_{j1} {V}_{k2}+ {\sum_{i,j=1}^3} C^{ij1l} {V}_{j1} {V}_{i2}\right),
\end{eqnarray}
where $\frac{{\alpha_2}}{2{\pi}}f_{\Delta}\sim \frac{{\alpha_2}}{2{\pi}}\frac{M_{wino}}{M_{SUSY}^2} \sim \frac{{\alpha_2}}{2{\pi}}\times2.5\times10^{-5}\textrm{GeV}^{-1}$  is the triangle diagram factor for $M_{wino}=400\textrm{GeV}$ and $M_{SUSY}=4\textrm{TeV}$.
$A_L=0.22$ is the long-distance renormalization factor \cite{Hisano Murayama}, and
$V_{ij}$s are the CKM matrix elements.
The short-distance renormalization factor $A_S$ is
\begin{equation}\label{pAs}
A_S=\left(\frac{\alpha_1(m_Z)}{\alpha_{10}(M_{GUT})}\right)^{-\frac{1}{33}} \left(\frac{\alpha_2(m_Z)}{\alpha_{10}(M_{GUT})}\right)^{-3} \left(\frac{\alpha_3(m_Z)}{\alpha_{10}(M_{GUT})}\right)^{\frac{4}{3}}\sim 7.16,
\end{equation}
if we take $\alpha_{10}(M_{GUT})=\frac{1}{25}$. The decay rates for $p\rightarrow K^+\nu_l$ are
\begin{eqnarray}
\Gamma(p\rightarrow K^+\nu_l)&=&\frac{{\beta_p}^2\left({m_p}^2-{m_K}^2\right)^2}{32 {\pi} {m_p}^3{f_\pi}^2} \left|\frac{2{m_p}}{3{m_B}}{D} C_{sudv_l}+(1+\frac{{m_p}}{3{m_B}}({D}+3{F}))C_{dusv_l}\right|^2,
\end{eqnarray}
where $m_p=0.938\textrm{GeV}$, $m_K=0.494\textrm{GeV}$, $f_\pi =0.131\textrm{GeV}$,
$D=0.81$, $F=0.44$ \cite{Hisano Murayama}, and $\beta_p=0.012\textrm{GeV}^3$ \cite{lattice}
are hadronic parameters.

To estimate proton decay rates, we need to analyze the effective triplet mass matrix $M_{eff}$ in
(\ref{effectivemass}) in some details. In getting the right-handed neutrino masses,
the VEV $\overline{v}_{R}$ is taken a small value $10^{14}$GeV compared to
the GUT scale $2\times 10^{16}$GeV.
In the limit of neglecting $\overline{v}_{R}$, there are three zero eigenvalues in the matrix
$B_{22}$, since in this limit equations (\ref{equphi2}--\ref{equphi3})
require $\eta_{11}/\eta_{21}\sim \eta_{12}/\eta_{22}$ and thus
$B_{22}$ has rank-4 instead of  rank-7. Consequently,
the elements in $M_{eff}$ are all divergent.
However, these elements are correlated, thus elements in the inverse of $M_{eff}$ are not small even in this limit. Proton decay proceeds fastly whenever the  eigenvalues in $M_{eff}$ are not all very big, being irrelevant to the appearance of all the large entries.

To suppress proton decay,
we need to keep all the eigenvalues in $B_{12}$ and $B_{21}$ at GUT scale while suppressing those in $B_{22}$.
The later can be achieved only by fine-tuning slightly the parameters $\eta_{ij} (i,j=1,2)$ and
$m_{\Phi}+\lambda(\frac{\Phi_1}{\sqrt{6}}+\frac{\Phi_2}{3\sqrt{2}}+\frac{2\Phi_3}{3})$.
The former, to keep the eigenvalues in $B_{12}$ and $B_{21}$ all  large,
requires at least one pair of large VEVs
in  $\overline{v}_{iR}$ (see the last row in the matrix of (\ref{MT12}))
and $v_{iR}$ (see the last column in the matrix of (\ref{MT21})).
These large $\overline{v}_{iR}$ and $v_{iR}$, now required by suppressing proton decay, correspond to the direct breaking of SO(10) into MSSM, and explicitly breaking of the unification is avoided as
all the Higgs particles beyond MSSM are at GUT scale.

Numerically, there are too many parameters in the model to analyze.
The typical dimensional and dimensionless  parameters are taken as in Table \ref{dimension1}
and \ref{dimensionless1}, respectively.
They lead to the VEVs in Table \ref{VEVs}.
\begin{table}[t]
\begin{center}
\begin{tabular}{|c|c|c|c|c|c|c|}
\hline
parameter & $m_{\Phi}$  & $m_{\Delta_1}$  & $m_{\Delta_2}$ & $\overline{m}_{\Delta_1}$& $\overline{m}_{\Delta_2}$ & $X$\\
\hline
value & $1$ &3 &2.96 &$5$ &$2$ & 30 \\
\hline
\end{tabular}
\caption{Dimensional parameters(in $10^{16}\textrm{GeV}$).}\label{dimension1}
\end{center}
\end{table}

\begin{table}[t]
\begin{center}
\begin{tabular}{|c|c|c|c|c|c|c|c|c|c|c|c|}
\hline
parameter & x &$\lambda$  & $\lambda_1$  & $\lambda_2$ & $\overline{\lambda}_1$& $\overline{\lambda}_2$& ${\gamma}_1$ & ${\gamma}_2$ & $\overline{\gamma}_1$ & $\overline{\gamma}_2$ & $\eta_{11}$  \\
\hline
value & 0.17 & $0.25$ & $0.6$ & $-0.62$ &$1.8$ &$1.32$  &$1.5$ &$1.28$ &$1.05$ &$0.89$ & $0.02$  \\
\hline
parameter & $\eta_{12}$ & $\eta_{21}$ & $\eta_{22}$& $\beta_{11}$& $\beta_{12}$ & $\beta_{21}$&$\beta_{22}$ & $\overline{\beta}_{11}$ & $\overline{\beta}_{12}$ & $\overline{\beta}_{21}$ & $\overline{\beta}_{22}$ \\
\hline
value & $0.015$ & $0.025$ &$0.0188$ &$-1$ &$1.2$ &$1.5$ &$1.35$ &$-1.5$ &$1.24$ &$1.35$ &$1.35$ \\
\hline
\end{tabular}
\caption{Dimensionless parameters. }\label{dimensionless1}
\end{center}
\end{table}

\begin{table}[!t]
\begin{center}
\begin{tabular}{|c|c|c|c|c|c|c|c|c|c|}
\hline
VEV & $\Phi_{1}$  & $\Phi_{2}$ & $\Phi_{3}$ & $v_{R}$& $\overline{v}_{R}$& $v_{1R}$& $\overline{v}_{1R}$&$v_{2R}$& $\overline{v}_{2R}$\\
\hline
value & $2.1$ & $12.9$ &$4.08$ &$1.82$ &$0$ &$5.66$ &$18.23$ &$15.0$ &$11.94$  \\
\hline
\end{tabular}
\caption{The VEVs(in $10^{16}\textrm{GeV}$).}\label{VEVs}
\end{center}
\end{table}
Note that the  VEVs $v_R$s and $\bar{v}_R$s are normalized to $\frac{1}{\sqrt{120}}$ \cite{tfukuyama},
thus the appearance of their large values is artificial. Indeed, all the gauge superfields get masses
at the GUT scale through the VEVs listed in Table \ref{VEVs}. Equations (\ref{dterm}--\ref{equX}) are satisfied. $\overline{v}_{R}=0$ is taken to study proton decay; in analyzing neutrino masses, however,
it should take its practical values like $10^{14}$GeV
which are still negligible compared to the GUT scale.

We have also checked numerically the entire Higgs spectrum,  confirming the absence of
intermediate state which may otherwise spoil the gauge coupling unification  explicitly.
There is the problem, however, that the total $\beta-$function will be a large number after unification, so the coupling constant of SO(10) will blow up quickly.

The numerical results of proton partial lifetimes are listed in Table \ref{lifetime}.
As we can see, they can be well above the current experimental limit,
although at the cost of fine-tuning some parameters.
According to the recent discussion of \cite{revivesu5}, by taking decoupling effects of SUSY particles into account, the  triangle diagram factors can be even smaller, leading to longer proton lifetime.
\begin{table}[t]
\begin{center}
\begin{tabular}{cl}
\hline
channel & \quad\quad\quad\quad\quad\quad\quad\quad\quad\quad\quad {lifetime} \\
\hline
$\tau(p\rightarrow K^+\nu_e)$& = {$3.81 \times 10^{36}$yrs } \\
$\tau(p\rightarrow K^+\nu_\mu)$ &= {$1.08 \times 10^{34}$yrs } $ \times \left|\frac{0.012\textrm{\scriptsize{GeV}}^3}{\beta_p} \left(\frac{\alpha_{10}^{-1}}{25}\right)^{\frac{56}{33}} \frac{2.5\times10^{-5}\textrm{\scriptsize{GeV}}^{-1}}{f_{\Delta}}\right|^2 $\\
$\tau(p\rightarrow K^+\nu_\tau)$& = {$1.60 \times 10^{34}$yrs } \\
\hline
\end{tabular}
\caption{Proton partial lifetime /years}\label{lifetime}
\end{center}
\end{table}

\section{Summary and conclusions}\label{six}

The extended SO10MPM is analyzed with the following results. First,
type-I see-saw can be realized by introducing an intermediate VEV which couples
to fermions. Second, SUSY is maintained at high energy. Third, unification is
hopefully realized although a fully adjustment of the parameters are not carried out.
Fourth, proton lifetime is consistent with the data if fine-tuning is used slightly.
Works we have not done here include a fully numerical calculation of
gauge coupling unification including GUT scale threshold effects, and a fully analysis
with electron mass corrected by the 120plet Higgs effect. These are big challenges
in future researches. We thank Jun-hui Zheng for useful discussions.



\begin{thebibliography}{99}

\bibitem{clark1982}
T.~E.~Clark, T.~K.~Kuo and N.~Nakagawa, \textit{An} SO(10) \textit{supersymmetric grand unified theory}, \textit{Phys.~Lett.}~\textbf{B~115}~(1982)~26.

\bibitem{aulakh1983}
C.~S.~Aulakh and R.~N.~Mohapatra, \textit{Implications of supersymmetric} SO(10) \textit{grand unification}, \textit{Phys.~Rev.}~\textbf{D~28}~(1983)~217.

\bibitem{babu1993}
K.~S.~Babu and R.~N.~Mohapatra, \textit{Predictive neutrino spectrum in minimal} SO(10) \textit{grand unification}, \textit{Phys. Rev. Lett.}~\textbf{70}~(1993)~2845.

\bibitem{bajc2003}
B.~Bajc, G.~Senjanovic and F.~Vissani, \textit{$b-\tau$ unification and large atmospheric mixing: a case for non-canonical see-saw}, \textit{Phys. Rev. Lett.}~\textbf{90}~(2003)~051802.

\bibitem{goh2003}
H.~S.~Goh, R.~N.~Mohapatra and S.-P.~Ng, \textit{Minimal SUSY} SO(10), \textit{$b-\tau$ unification and large neutrino mixings}, \textit{Phys. Lett.}~\textbf{B~570}~(2003)~215. 

\bibitem{aulakh2004}
C.~S.~Aulakh, B.~Bajc, A.~Melfo, G.~Senjanovic and F.~Vissani, \textit{The minimal supersymmetric grand unified theory}, \textit{Phys.~Lett.}~\textbf{B~588}~(2004)~196.

\bibitem{fukuyama2004}
T.~Fukuyama, T.~Kikuchi, A.~Ilakovac, S.~Meljanac and N.~Okada, \textit{Detailed analysis of proton decay rate in the minimal supersymmetric} SO(10) \textit{model}, \textit{JHEP}~\textbf{0409}~(2004)~052.

\bibitem{bajc2004}
B.~Bajc, A.~Melfo, G.~Senjanovi$\acute{\textrm{c}}$ and F.~Vissani, \textit{Minimal supersymmetric grand unified theory: Symmetry breaking and the particle spectrum}, \textit{Phys.~Rev.}~\textbf{D~70}~(2004)~035007.

\bibitem{fukuyama2005}
T.~Fukuyama, A.~Ilakovac, T.~Kikuchi, S.~Meljanac and N.~Okada, \textit{General formulation for proton decay rate in minimal supersymmetric} SO(10) \textit{GUT}, \textit{Eur.Phys.J.}~\textbf{C~42}~(2005)~191.

\bibitem{fukuyama2005b}
T.~Fukuyama, A.~Ilakovac, T.~Kikuchi, S.~Meljanac and N.~Okada, \textit{Higgs Masses in the Minimal SUSY} SO(10) \textit{GUT},
\textit{Phys.~Rev.}~\textbf{D~72}~(2005)~051701.

\bibitem{babu2005}
K.~S.~Babu and C.~Macesanu, \textit{Neutrino masses and mixings in a minimal} SO(10) \textit{model}, \textit{Phys. Rev.}~\textbf{D~72}~(2005)~115003.

\bibitem{aulakh2005}
C.~S.~Aulakh and A.~Girdhar, SO(10) \textit{MSGUT: Spectra, couplings and threshold effects}, \textit{Nucl.~Phys.}~\textbf{B~711}~(2005)~275.

\bibitem{bertolini2006}
S.~Bertolini, T.~Schwetz and M.~Malinsky, \textit{Fermion masses and mixings in} SO(10) \textit{models and the neutrino challenge to supersymmetric grand unified theories}, \textit{Phys. Rev.}~\textbf{D~73}~(2006)~115012.

\bibitem{aulakh2006}
C.~S.~Aulakh and S.~K.~Garg, \textit{MSGUT: From bloom to doom}, \textit{Nucl.~Phys.}~\textbf{B~757}~(2006)~47.

\bibitem{dutta2008}
B.~Dutta, Y.~Mimura and R.~N.~Mohapatra, \textit{Proton decay and flavor violating thresholds in} SO(10) \textit{models}, \textit{Phys. Rev. Lett.}~\textbf{100}~(2008)~181801.

\bibitem{aulakh2009}
C.~S.~Aulakh and S.~K.~Garg, \textit{Correcting $\alpha_3(M_Z)$ in the NMSGUT}, \textit{Mod. Phys. Lett.}~\textbf{A~24}~(2009)~1711.

\bibitem{aulakh2012}
C.~S.~Aulakh and S.~K.~Garg, \textit{The new minimal supersymmetric GUT: spectra, RG analysis and fermion fits}, \textit{Nucl.~Phys.}~\textbf{B~857}~(2012)~101.

\bibitem{aulakh2013}
C.~S.~Aulakh, I.~Garg and C.~K.~Khosa, \textit{Baryon stability on the Higgs dissolution edge: threshold corrections and suppression of baryon violation in the NMSGUT} [arXiv:hep-ph/1311.6100].

\bibitem{MPM masiero1982}
A.~Masiero, D.~V.~Nanopoulos, K.~Tamvakis and T.~Yanagida, \textit{Naturally massless Higgs doublets in supersymmetric} SU(5), \textit{Phys.~Lett.}~\textbf{B~115}~(1982)~380.

\bibitem{MPM grinstein1982}
B.~Grinstein, \textit{A supersymmetric} SU(5) \textit{gauge theory with no gauge hierarchy problem}, \textit{Nucl.~Phys.}~\textbf{B~206}~(1982)~387.

\bibitem{MPM hisano1995}
J.~Hisano, T.~Moroi, K.~Tobe and T.~Yanagida, \textit{Suppression of proton decay in the missing partner model for supersymmetric} SU(5) \textit{GUT}, \textit{Phys.~Lett.}~\textbf{B~342}~(1995)~138.

\bibitem{MPM zheng2012}
J.~Zheng and D.-X.~Zhang, \textit{A renormalizable supersymmetric} SU(5) \textit{model}, \textit{JHEP}~\textbf{1202}~(2012)~046.

\bibitem{MPM zhang2012}
D.-X.~Zhang and J.~Zheng, \textit{A missing partner model with} 24\textit{-plet breaking} SU(5), \textit{JHEP}~\textbf{1212}~(2012)~087.

\bibitem{pulido1983}
J.~Maalampi and J.~Pulido, \textit{Locally supersymmetric} SO(10) \textit{with natural doublet/triplet splitting},
\textit{Phys.Lett.}~\textbf{B~133}~(1983)~197.

\bibitem{ksbabu}
K.~S.~Babu, I.~Gogoladze and Z.~Tavartkiladze, \textit{Missing partner mechanism in} SO(10) \textit{grand unification}, \textit{Phys.~Lett.}~\textbf{B~650}~(2007)~49.

\bibitem{babu2}
K.~S.~Babu, I.~Gogoladze, P.~Nath and  R.~M.~Syed, \textit{Variety of} SO(10) \textit{GUTs with natural doublet-triplet splitting via the missing partner mechanism}, \textit{Phys.~Rev.}~\textbf{D~85}~(2012)~075002.

\bibitem{typeone1}
P.~Minkowski, $\mu\to e\gamma$ \textit{at a rate of one out of 109 muon decays}?, \textit{Phys.~Lett.}~\textbf{B~67}~(1977)~421.

\bibitem{typeone2}
T.~Yanagida, in \textit{workshop on unified theories}, KEK Report 79-18~(1979)~95.

\bibitem{typeone3}
M.~Gell-Mann, P.~Ramond and R.~Slansky,
in  \textit{Supergravity}, P. van Nieuwenhuizen and D.Z. Freedman (eds.), North Holland Publ. Co., 1979

\bibitem{typeone4}
R.~N.~Mohapatra and G.~Senjanovi$\acute{\textrm{c}}$, \textit{Neutrino mass and spontaneous parity nonconservation}, \textit{Phys.~Rev.~Lett.}~\textbf{44}~(1980)~912.

\bibitem{nurscale2002}
S.~DavidsonA and A.~Ibarra, \textit{A lower bound on the right-handed neutrino mass from leptogenesis}, \textit{Phys.~Lett.}~\textbf{B~535}~(2002)~25.

\bibitem{nurscale2004}
C.~H.~Albright, \textit{Normal vs. inverted hierarchy in type I seesaw models}, \textit{Phys. Lett.}~\textbf{B~599}~(2004)~285.

\bibitem{nurscale2013}
D.~Borah and M.~K.~Das, \textit{Neutrino masses and mixings with non-zero $\theta_{13}$ in Type I+II Seesaw Models}, \textit{Nucl.~Phys.}~\textbf{B~870}~(2013)~461.

\bibitem{green1984}
M.~B.~Green and J.~H.~Schwarz, \textit{Anomaly cancellations in supersymmetric $D=10$ gauge theory and superstring theory}, \textit{Phys. Lett.}~\textbf{B~149}~(1984)~117.

\bibitem{dine1987}
M.~Dine, N.~Seiberg and E.~Witten, \textit{Fayet-Iliopoulos terms in string theory}, \textit{Nucl. Phys.}~\textbf{B~289}~(1987)~589.

\bibitem{atick1987}
J.~J.~Atick, L.~J.~Dixon and A.~Sen, \textit{String calculation of fayet-iliopoulos D-terms in arbitrary supersymmetric compactifications}, \textit{Nucl. Phys.}~\textbf{B~292}~(1987)~109.

\bibitem{dine19872}
M.~Dine, I.~Ichinose and N.~Seiberg, \textit{F terms and D terms in string theory}, \textit{Nucl. Phys.}~\textbf{B~293}~(1987)~253.

\bibitem{tfukuyama}
T.~Fukuyama, A.~Ilakovac, T.~Kikuchi, S.~Meljanac and N.~Okada, SO(10) \textit{Group theory for the unified model building}, \textit{J.~Math.~Phys.} \textbf{46}~(2005)~033505.

\bibitem{seesawII}
B.~Bajc, G.~Senjanovi$\acute{\textrm{c}}$ and F.~Vissani, \textit{Probing the nature of the seesaw in renormalizable} SO(10), \textit{Phys.~Rev.}~\textbf{D~70}~(2004)~093002.

\bibitem{bertolini}
S.~Bertolini, M.~Frigerio and M.~Malinsk$\acute{\textrm{y}}$, \textit{Fermion masses in supersymmetric} SO(10) \textit{with type II seesaw mechanism: A nonminimal predictive scenario}, \textit{Phys.~Rev.}~\textbf{D~70}~(2004)~095002.

\bibitem{Patel}
A.~S.~Joshipura and K.~M.~Patel, \textit{Fermion masses in} SO(10) \textit{models}, \textit{Phys.~Rev.}~\textbf{D~83}~(2011)~095002.

\bibitem{C.R.Das}
C.~R.~Das and M.~K.~Parida, \textit{New formulas and predictions for running fermion masses at higher scales in SM, 2HDM, and MSSM},
\textit{Eur.~Phys.~J.}~\textbf{C~20}~(2001)~121.

\bibitem{nath1985}
P.~Nath, A.~H.~Chamseddine and R.~L.~Arnowitt, \textit{Nucleon Decay in Supergravity Unified Theories}, \textit{Phys. Rev.}~\textbf{D~32}~(1985)~2348.

\bibitem{goto1999}
T.~Goto and T.~Nihei, \textit{Effect of RRRR dimention five operator on the proton decay in the minimal} SU(5) \textit{SUGRA GUT model}, \textit{Phys.~Rev.}~\textbf{D~59}~(1999)~115009.

\bibitem{Hisano Murayama}
J.~Hisano, H.~Murayama and T.~Yanagida, \textit{Nucleon decay in the minimal supersymmetric} SU(5) \textit{grand unification}, \textit{Nucl.~Phys.}~\textbf{B~402}~(1993)~46.

\bibitem{lattice}
Y.~Aoki, P.~Boyle, P.~Cooney, L.~D.~Debbio, R.~Kenway, C.~M.~Maynard, A.~Soni and R.~Tweedie, \textit{Proton lifetime bounds from chirally symmetric lattice QCD}, \textit{Phys.~Rev.}~\textbf{D~78}~(2008)~0545505.

\bibitem{revivesu5}
J.~Hisano, D.~Kobayashi, T.~Kuwahara and N.~Nagata, \textit{Decoupling can revive minimal supersymmetric} SU(5), \textit{JHEP}~\textbf{1307}~(2013)~038.


\end{thebibliography}
\end{document}